\documentstyle[12pt]{article}
\begin{document}
\centerline{\Large\bf  Black hole and cosmological space-times in}
\vskip 0.1in
\centerline{\Large\bf Born-Infeld-Einstein theory}
\vskip .7in
\centerline{Dan N. Vollick}
\centerline{Irving K. Barber School of Arts and Sciences}
\centerline{University of British Columbia Okanagan}
\centerline{3333 University Way}
\centerline{Kelowna, B.C.}
\centerline{V1V 1V7}
\vskip .9in
\centerline{\bf\large Abstract}
\vskip 0.5in
In this paper I examine black hole and cosmological space-times in Born-Infeld-Einstein
theory with electric and magnetic charges. The field equations are derived
and 
written in the form $G_{\mu\nu}=-\kappa T_{\mu\nu}$ for spherically symmetric
space-times. The energy-momentum tensor is not the Born-Infeld energy-momentum
tensor, but can be obtained from Born-Infeld theory by letting $a\rightarrow
ia$, where $a$ is the Born-Infeld parameter. It is shown that there is a
curvature singularity in spherically symmetric space-times  at a nonzero
radial coordinate and that, as in Reissner-Nordstrom space-times, there
are zero, one or two horizons. Charged black holes have either two horizons
and a timelike singularity or one horizon with a spacelike, timelike, or
null singularity.
Anisotropic cosmological solutions with electric
and magnetic fields are obtained
from the spherically symmetric solutions. 
\newpage
\section*{Introduction}
Born-Infeld electrodynamics follows from the Lagrangian \cite{Bo1}
\begin{equation}
L=-\frac{1}{4\pi b}\left\{\sqrt{-det(g_{\mu\nu}+bF_{\mu\nu})}
-\sqrt{-det(g_{\mu\nu})}\right\}\; ,
\end{equation}
where $g_{\mu\nu}$ is the metric tensor and $F_{\mu\nu}$ is the
electromagnetic field tensor.
In the weak field limit this Lagrangian reduces to the
Maxwell Lagrangian plus small corrections. For strong fields the
field equations deviate significantly from Maxwell's theory and 
the self energy of the electron can be shown to be finite.
The Born-Infeld action also appears in string theory. The action
for a D-brane is of the Born-Infeld form with two fields, a
gauge field on the brane and the projection of the Neveu-Schwarz
B-field onto the brane \cite{Po1}.
   
In recent papers \cite{Vo1,Vo2} I considered using a Palatini variational
approach to derive the field equations associated with the Lagrangian
\begin{equation}
L=-\frac{1}{\kappa b}\left\{\sqrt{-det(g_{\mu\nu}+bR_{\mu\nu}+\kappa bM_{\mu\nu})}
-\sqrt{-det(g_{\mu\nu})}\right\},
\label{intro}
\end{equation}
where $R_{\mu\nu}$ is the Ricci tensor, $M_{\mu\nu}$ is the matter contribution,
and $\kappa=8\pi G$. This Lagrangian, with $M_{\mu\nu}=0$,
has also been examined using a purely metric
variation by Deser and Gibbons \cite{De1}, Feigenbaum, Freund and Pigli 
\cite{Fe1} and Feigenbaum \cite{Fe2}.

In this paper I examine black hole and cosmological space-times in Born-Infeld-Einstein
theory with electric and magnetic charges. The field equations are derived
for spherically symmetric space-times and it is shown that they can be written
in the form $G_{\mu\nu}=-\kappa T_{\mu\nu}$. The energy-momentum tensor is not the Born-Infeld energy-momentum
tensor, but can be obtained from Born-Infeld theory by letting $a\rightarrow
ia$, where $a$ is the Born-Infeld parameter. The field equations are solved
and it is shown that the space-time has a curvature singularity at a nonzero
radial coordinate and that, as in Reissner-Nordstrom space-times, there
are zero, one or two horizons. Thus, black holes in this theory will have
one or two horizons. 
Anisotropic cosmological solutions with electric and magnetic fields 
are obtained from the spherically symmetric solutions. 
 
\section*{The Field Equations}
In this section I will derive the field equations for the Born-Infeld-Einstein
theory with an electromagnetic field and a cosmological constant. 
The action is given by
\begin{equation}
L=-\frac{1}{\kappa b}\left\{\sqrt{-det(g_{\mu\nu}+
bR_{\mu\nu}+\sqrt{\kappa b}F_{\mu\nu}+b\lambda g_{\mu\nu})}-\sqrt{-det(g_{\mu\nu})}\right\}\;,
\label{lag}
\end{equation}
where $R_{\mu\nu}$ is the Ricci tensor, $\kappa=8\pi G$, $b$ 
is a constant, $F_{\mu\nu}$ is the electromagnetic field tensor and $\lambda$
is a constant.
The Ricci tensor is given by
\begin{equation}
R_{\mu\nu}=\partial_{\nu}\Gamma^{\alpha}_{\mu\alpha}-\partial_{\alpha}
\Gamma^{\alpha}_{\mu\nu}-\Gamma^{\alpha}_{\beta\alpha}\Gamma^{\beta}_{\mu\nu}
+\Gamma^{\alpha}_{\beta\mu}\Gamma^{\beta}_{\alpha\nu}
\label{Ricci1}
\end{equation}
and the connection is taken to be symmetric. Note that $R_{\mu\nu}$ is not
symmetric in general. The electromagnetic field term appears in the action
multiplied by the constant $(\kappa b)^{1/2}$ instead of $\kappa b$ since
the lowest order term in the expansion of the Lagrangian is quadratic in
$F_{\mu\nu}$.

Varying the action with respect to $g_{\mu\nu}$ gives
\begin{equation}
(1+b\lambda)\sqrt{P}\left(P^{-1}\right)^{(\mu\nu)}=\sqrt{g}g^{\mu\nu}
\label{eq1}\; ,
\end{equation}
where $P_{\mu\nu}=(1+b\lambda)g_{\mu\nu}+bR_{\mu\nu}+(\kappa b)^{1/2}F_{\mu\nu}$, $P^{-1}$ is the inverse of $P$, $(P^{-1})^{(\mu\nu)}$
is the symmetric part of $P^{-1}$, $P=-det(P_{\mu\nu})$ and $g=-det(g_{\mu\nu})$.
  
In the above action I have included the cosmological constant term 
in the first determinant along with the electromagnetic source term. However,
the cosmological term could be added to the second determinant instead giving
$-\sqrt{-det((1+b\lambda)g_{\mu\nu})}$ for this term. 
The resulting field equations, written in terms of $\bar{g}_{\mu\nu}$
and $\bar{\lambda}$ can
be obtained from the above field equations by making the following substitutions
\begin{equation}
g_{\mu\nu}=(1+b\bar{\lambda})\bar{g}_{\mu\nu}\;\;\;\;\; and \;\;\;\;\;
1+b\lambda=\frac{1}{{1+b\bar{\lambda}}}\;.
\end{equation}
A cosmological term could also be added to both terms in various ways, but
I will not consider these possibilities here.

Varying with the action with respect to $\Gamma^{\alpha}_{\mu\nu}$ gives
\begin{equation}
\nabla_{\alpha}\left[\sqrt{P}\left(P^{-1}\right)^{(\mu\nu)}\right]
-\frac{1}{2}\nabla_{\beta}\left\{\sqrt{P}\left[\delta^{\mu}_{\alpha}
\left(P^{-1}\right)^{\beta\nu}+\delta^{\nu}_{\alpha}\left(P^{-1}
\right)^{\beta\mu}\right]\right\}=0\; .
\label{eq2}
\end{equation}
In an earlier paper \cite{Vo1} I showed that this equation together with the electromagnetic
field equation, which will be discussed next, implies that the connection is the
Christoffel symbol.
  
The field equations for the electromagnetic field are derived by varying
the action with respect to $A_{\mu}$ and are given by
\begin{equation}
\nabla_{\mu}\left[\sqrt{P}\left(P^{-1}\right)^{[\mu\nu]}\right]=0\;.
\label{eq22a}
\end{equation}
  
The above field equations are difficult to solve in general. 
However, it is possible to solve
them for spherically symmetric electric and magnetic fields. 
In this case
\begin{equation}
P_{\mu\nu}=\left[
\begin{array}{ll}
\;\;q_{00}\;\;\;\;\;\;E'\;\;\;\;\;0\;\;\;\;\;\;0\\
\;-E'\;\;\;\;\;q_{11}\;\;\;\;0\;\;\;\;\;\;0\\
\;\;\;0\;\;\;\;\;\;\;\;0\;\;\;\;\;q_{22}\;\;\;B'\\
\;\;\;0\;\;\;\;\;\;\;\;0\;\;\;-B'\;\;q_{33}\\
\end{array}
\right]
\label{P}
\end{equation}
and
\begin{equation}
\left(P^{-1}\right)^{\mu\nu}=\left[
\begin{array}{ll}
\;\frac{q_{11}}{\Delta_1}\;\;-\frac{E'}{\Delta_1}\;\;\;\;\;\;0\;\;\;\;\;\;0\\
\\
\;\frac{E'}{\Delta_1}\;\;\;\;\;\;\;\frac{q_{00}}{\Delta_1}\;\;\;\;\;0\;\;\;\;\;\;0\\
\\
\;\;0\;\;\;\;\;\;\;\;\;0\;\;\;\;\;\frac{q_{33}}{\Delta_2}\;\;\;\;\;\frac{-B'}{\Delta_2}\\
\\
\;\;0\;\;\;\;\;\;\;\;\;0\;\;\;\;\;\;\frac{B'}{\Delta_2}\;\;\;\;\;\frac{q_{22}}{\Delta_2}\\
\end{array}
\right]\; ,
\end{equation}
where $q_{\mu\nu}=(1+b\lambda)g_{\mu\nu}+bR_{\mu\nu}$, 
$E'=\sqrt{\kappa b}E$, $B'=\sqrt{\kappa b}B$, 
$\Delta_1=q_{00}q_{11}+
E^{'2}$ and $\Delta_2=q_{22}q_{33}+B^{'2}$.
If $\Delta_1<0$ and $\Delta_2>0$ at some point in space-time then
this must be true everywhere, if $\Delta_1$ and $\Delta_2$ are continuous,
since $P=-\Delta_1\Delta_2\neq 0$.
This will be the case for
the space-times examined in this paper and I will therefore take $\Delta_1<0$ and $\Delta_2>0$.
  
Taking the determinant of both sides of
(\ref{eq1}) gives
\begin{equation}
g=-(1+b\lambda)^4q_{00}q_{11}q_{22}q_{33}\; .
\end{equation}
The field equations can then be written as
\begin{equation}
\left(P^{-1}\right)^{(\mu\nu)}=(1+b\lambda)\sqrt{\frac{q_{00}q_{11}q_{22}q_{33}}
{\Delta_1\Delta_2}}\;g^{\mu\nu}\;
\end{equation}
and give the following four equations
\begin{equation}
q_{00}=-(1+b\lambda)\sqrt{\frac{q_{00}q_{11}q_{22}q_{33}\Delta_1}{\Delta_2}}
\;\;g^{11}\; ,
\label{eq00}
\end{equation}
\begin{equation}
q_{11}=-(1+b\lambda)\sqrt{\frac{q_{00}q_{11}q_{22}q_{33}\Delta_1}{\Delta_2}}
\;\;g^{00}\; ,
\label{eq11}
\end{equation}
\begin{equation}
q_{22}=(1+b\lambda)\sqrt{\frac{q_{00}q_{11}q_{22}q_{33}\Delta_2}{\Delta_1}}
\;\;g^{33}\; ,
\label{eq22}
\end{equation}
and 
\begin{equation}
q_{33}=(1+b\lambda)\sqrt{\frac{q_{00}q_{11}q_{22}q_{33}\Delta_2}{\Delta_1}}
\;\;g^{22}\; .
\label{eq33}
\end{equation}
From (\ref{eq00}) and (\ref{eq11}) we find that
\begin{equation}
(1+b\lambda)^2q_{22}q_{33}\Delta_1=g_{00}g_{11}\Delta_2\;\;\;\;\;\; and\;\;\;\;\;
g_{00}R_{11}=g_{11}R_{00}
\end{equation}
and from (\ref{eq22}) and (\ref{eq33}) we find that
\begin{equation}
(1+b\lambda)^2q_{00}q_{11}\Delta_2=g_{22}g_{33}\Delta_1\;\;\;\;\;\; and\;\;\;\;\;
g_{22}R_{33}=g_{33}R_{22}\;.
\end{equation}
Using these equations gives 
\begin{equation}
g_{11}b^2R_{00}^2+2(1+b\lambda)g_{00}g_{11}bR_{00}+\frac{g_{00}E^{'2}-g_{00}^2g_{11}
[(1+b\lambda)^{-2}-(1+b\lambda)^2(1-a^2\tilde{B}^2)]}{(1-a^2\tilde{B}^2)}=0\;,
\end{equation}
where
\begin{equation}
\tilde{B}^2=(1+b\lambda)^2g^{22}g^{33}B^2\; .
\end{equation}
and $\kappa b=a^2$.
The solution to this equation is
\begin{equation}
R_{00}=-\frac{g_{00}}{b}\left[(1+b\lambda)-(1+b\lambda)^{-1}\sqrt{\frac{1+a^2\tilde{E}^2}
{1-a^2\tilde{B}^2}}
\;\right]\;,
\label{R00}
\end{equation}
where $\tilde{E}^2=-(1+b\lambda)^2g^{00}g^{11}E^2$.
To find $R_{11}$ we can use $R_{11}=g_{11}R_{00}/g_{00}$. A calculation similar
to the one above gives
\begin{equation}
R_{22}=-\frac{g_{22}}{b}\left[(1+b\lambda)-(1+b\lambda)^{-1}\sqrt{\frac{1-a^2\tilde{B}^2}
{1+a^2\tilde{E}^2}}
\;\right]\;,
\label{R22}
\end{equation}
and $R_{33}$ is given by $R_{33}=g_{33}R_{22}/g_{22}$.
The solution can be expressed in terms of $\tilde{F}_{\mu\nu}=
(1+b\lambda)F_{\mu\nu}$ and the invariants $\tilde{F}^2=
\tilde{F}^{\mu\nu}\tilde{F}_{\mu\nu}$, $\tilde{S}^2=
\frac{1}{4}\star\tilde{F}^{\mu\nu}\tilde{F}_{\mu\nu}$ and
$\star\tilde{F}^{\mu\nu}=\frac{1}{2}\epsilon^{\mu\nu\alpha\beta}
\tilde{F}_{\alpha\beta}$. 
It is given by
\begin{equation}
G_{\mu\nu}=-(1+\tilde{\lambda})^{-1}\kappa \left\{\frac{\tilde{F}_{\mu}^{\;\;\;\alpha}
\tilde{F}_{\nu\alpha}}{\sqrt{
1-\frac{1}{2}a^2\tilde{F}^2-a^4\tilde{S}^4}}-\frac{g_{\mu\nu}}{a^2}
\left[(1+\tilde{\lambda})^2-\frac{1-\frac{1}{2}a^2\tilde{F}^2}
{\sqrt{1-\frac{1}{2}a^2\tilde{F}^2-a^4\tilde{S}^4}}\;\right]\right\}\;,
\label{EFE}
\end{equation}
where $\tilde{\lambda}=a^2\lambda/\kappa$.
Note that we require $1+\tilde{\lambda}>0$ so that the effective Einstein
constant $(1+\tilde{\lambda})^{-1}\kappa$ is greater than zero.
Also note that if $\lambda=0$
the energy-momentum tensor is identical
to the Born-Infeld energy-momentum tensor, except for the sign in front of
the $\tilde{F}^2$ term  in the square roots and the sign in front of 
the metric. Thus, the Einstein field equations in the absence of a cosmological
constant can be obtained from the Born-Infeld theory by replacing $a$ 
by $ia$. The energy-momentum tensor that appears on the right hand side
of equation (\ref{EFE}) can be separated into a cosmological constant term
plus a contribution that vanishes if $F^{\mu\nu}=0$:
\begin{equation}
T_{\mu\nu}=-\frac{\Lambda}{\kappa}g_{\mu\nu}+(1+\tilde{\lambda})^{-1}
\left\{\frac{\tilde{F}_{\mu}^{\;\;\;\alpha}
\tilde{F}_{\nu\alpha}}{\sqrt{
1-\frac{1}{2}a^2\tilde{F}^2-a^4\tilde{S}^4}}-\frac{g_{\mu\nu}}{a^2}
\left[1-\frac{1-a^2\tilde{F}^2}
{\sqrt{1-\frac{1}{2}a^2\tilde{F}^2-a^4\tilde{S}^4}}\;\right]\right\}\;,
\label{energy}
\end{equation}
where
\begin{equation}
\Lambda=\lambda\left[\frac{2+b\lambda}{1+b\lambda}\right]\; .
\label{Lambda}
\end{equation}

The electromagnetic field equations given in (\ref{eq22a}) can be written
as 
\begin{equation}
\nabla_{\mu}\left\{\frac{\tilde{F}^{\mu\nu}+a^2\tilde{S}^2\star\tilde{F}^{\mu\nu}}
{\sqrt{1-\frac{1}{2}a^2\tilde{F}^2-a^4\tilde{S}^4}}\right\}=0\; .
\label{EMFE}
\end{equation}
Once again, in the absence of a cosmological constant these equations follow
from Born-Infeld theory by replacing $a$ by $ia$. 

In flat space-time with $B=0$ and $\lambda=0$
the square root in Born-Infeld theory 
is given by $\sqrt{1-a^2E^2}$, so that the maximum magnitude of the electric
field is $1/a$. In the theory presented here, with $\lambda=B=0$, 
the square root is $\sqrt{1+a^2E^2}$,
so that the square root does not, by itself, constrain the magnitude of 
the electric field (see \cite{Vo1} for a more detailed discussion).The situation
is reversed if $E=0$ and $B\neq 0$.

The electromagnetic field equations (\ref{EMFE}) and the electromagnetic
contribution to the energy momentum tensor (\ref{energy}) can be derived
from the Lagrangian
\begin{equation}
L=\frac{1}{a^2}(1+\tilde{\lambda})^{-1}\left[\sqrt{1-\frac{1}{2}a^2
\tilde{F}^2-a^4\tilde{S}^4}-1\right]\; .
\end{equation}
It is convenient to introduce the field $\tilde{D}^{\mu\nu}$ defined by
\begin{equation}
\tilde{D}^{\mu\nu}=-2\frac{\partial L}{\partial F_{\mu\nu}}\;.
\end{equation}
It is given by
\begin{equation}
\tilde{D}^{\mu\nu}=\frac{\tilde{F}^{\mu\nu}+a^2\tilde{S}^2\star\tilde{F}^{\mu\nu}}
{\sqrt{1-\frac{1}{2}a^2\tilde{F}^2-a^4\tilde{S}^4}}\;.
\label{D}
\end{equation}
The field equations (\ref{EMFE}) can be written as
\begin{equation}
\nabla_{\mu}\tilde{D}^{\mu\nu}=0\; 
\end{equation}
and the electric induction is defined to be $\tilde{D}_k=\tilde{D}_{k0}$.
\section*{Black hole space-times}
Now consider spherically symmetric black hole solutions. 
The general spherically symmetric solution for any theory with a 
Lagrangian of the form
\begin{equation}
-\frac{1}{2\kappa}R+L(F^2,S^2)\; ,
\end{equation}
is given by \cite{Ho1,Ho2,Pe1,Dem1,Wi1,Ol1} 
\begin{equation}
ds^2=-\left[1-\frac{2m(r)}{r}\right]dt^2+\left[1-\frac{2m(r)}{r}\right]^{-1}dr^2
+r^2d\Omega^2\;,
\label{metric}
\end{equation} 
\begin{equation}
D=\frac{Q}{r^2}\;dt\wedge dr\;\;\;\;\; and\;\;\;\;\; B=P\sin\theta\;d\theta\wedge
d\phi\; ,
\end{equation}
where
\begin{equation}
\frac{dm(r)}{dr}=\frac{1}{2}\Lambda r^2+4\pi r^2H(r)\; ,
\label{BI}
\end{equation}
and H is given by
\begin{equation}
H=(1+\tilde{\lambda})^{-1}\tilde{\textbf{E}}\cdot\tilde{\textbf{D}}-L\;.
\label{Ham}
\end{equation}
From (\ref{D}) and (\ref{Ham}) it is easy to show that
\begin{equation}
H=\frac{1}{a^2}(1+\tilde{\lambda})^{-1}\left[1-\sqrt{1-\frac{a^2\tilde{Z}^2}{r^4}}\right].
\end{equation}
and equation (\ref{BI}) can be written as
\begin{equation}
\frac{dm(r)}{dr}=\frac{1}{2}\Lambda r^2+\frac{4\pi}{a^2}(1+\tilde{\lambda})^{-1}\left[
r^2-\sqrt{r^4-a^2\tilde{Z}^2}\right]\;,
\label{mprime}
\end{equation}
where $\tilde{Z}^2=\tilde{Q}^2+\tilde{P}^2$,
$\tilde{Q}=(1+\tilde{\lambda})Q$ and $\tilde{P}=(1+\tilde{\lambda})P$.
Thus, $r\geq \sqrt{a\tilde{Z}}$. 
  
To see if the space-time becomes singular $r\rightarrow\sqrt{a
\tilde{Z}}$
consider the Ricci tensor which is given by
\begin{equation}
R=-2\left[\frac{rm^{''}+2m^{'}}{r^2}\right]\;.
\end{equation}
From (\ref{mprime}) we have
\begin{equation}
R=-4\Lambda-\frac{32\pi}{a^2}(1+\tilde{\lambda})^{-1}\left[1-
\frac{r^2-\frac{a^2\tilde{Z}^2}{2r^2}}{\sqrt{r^4-a^2
\tilde{Z}^2}}\right]\;.
\end{equation}
Thus, there is a curvature singularity at $r=\sqrt{a
\tilde{Z}}$.
  
To understand the horizon structure of a charged black hole we need to look for zeros of $1-2m(r)/r$. From (\ref{mprime}) we see that
\begin{equation}
\frac{2m(r)}{r}=\frac{2m_0}{r}+\frac{\Lambda}{3}r^2-\frac{8\pi}{a^2r}(a
\tilde{Z})^{3/2}(1+\tilde{\lambda})^{-1}
\int_{\frac{r}{\sqrt{a\tilde{Z}}}}^{\infty}\left[u^2-\sqrt{u^4-1}\right]du\;,
\label{m}
\end{equation}
where $m_0$ is a constant. The second term 
is the usual cosmological constant term and $m_0$ is the mass of the 
black hole.
The last term in this expression drops off as $1/r^2$ at large $r$ and is
related to the electric and magnetic charges of the black hole. In fact,
for $r>>\sqrt{aZ}$
\begin{equation}
\frac{8\pi}{a^2r}(a\tilde{Z})^{3/2}\left(1+\tilde{\lambda}
\right)^{-1}\int_{\frac{r}{\sqrt{a\tilde{Z}}}}^{\infty}\left[u^2-\sqrt
{u^4-1}\right]du\simeq \frac{4\pi(1+\tilde{\lambda})^{-1}
(\tilde{Q}^2+\tilde{P}^2)}{r^2}\; .
\end{equation}
Thus, the electric and magnetic charges of the black hole are given by
\begin{equation}
Q_{BH}=\sqrt{1+\tilde{\lambda}}\;Q \;\;\;\; and \;\;\;\;P_{BH}=\sqrt{1+\tilde{
\lambda}}\;P\; .
\end{equation}
The zeros of $g_{tt}$ can be found by solving the equation
\begin{equation}
h(r)=r-2m_0+\frac{8\pi}{a^2}\left(a
Z\right)^{3/2}
\int_{\frac{r}{\sqrt{aZ}}}^{\infty}\left[u^2-\sqrt{u^4-1}
\right]du=0\;,
\end{equation}
where I have taken $\Lambda=0$ for simplicity. First find the critical points
of $h(r)$. The derivative of $h(r)$ is given by
\begin{equation}
h^{'}(r)=1-\frac{8\pi}{a^2}[r^2-\sqrt{r^4-a^2Z^2}\;]\;.
\end{equation}
The solution to $h^{'}(r)=0$ for $r>\sqrt{aZ}$ is given by
\begin{equation}
r=\sqrt{\frac{a^2+64\pi^2Z^2}{16\pi}}\;\;\;\;\;\; if \;\;\;\;\;\; Z>
\frac{a}{8\pi}\;.
\end{equation}
If $Z<a/8\pi$ there is no solution. 
Thus, if $Z>a/8\pi$ the function
$h(r)$ has one local minimum (one can show that $h^{''}>0$ at the critical point)
and if $Z<a/8\pi$ it has no local maxima or minima. Note that as $a\rightarrow
0$ we obtain the Reissner-Nordstrom result $r=\sqrt{4\pi}Z$ for the minimum
of $rg_{tt}$.
Next consider $h(r)$ evaluated at the endpoints of the interval it is defined
on:
\begin{equation}
h(\sqrt{aZ})=\sqrt{aZ}-2m_0+\frac{8\pi}{a^2}(aZ)^{3/2}
\int_1^{\infty}\left[u^2-\sqrt{u^4-1}\right]du
\end{equation}
and
\begin{equation}
h(r)\rightarrow\infty\;\;\;\; as\;\;\;\; r\rightarrow\infty\; .
\end{equation}
Now consider fixing the electric and magnetic charges of the black hole and
letting its mass vary. If $m_0$ is large enough $h(\sqrt{aZ})$ will be negative
and there will be only one horizon. In this case the singularity will be spacelike.
If $m_0$ is sufficiently
small, so that $h(\sqrt{aZ})\geq 0$, there can be zero, one or two 
horizons and the singularity will be timelike in all three cases. 
If $h(\sqrt{aZ})=0$ the singularity is null and there will be either zero
or one horizon outside the singularity.
It is interesting to note that this horizon structure
is the same as in the Reissner-Nordstrom space-time if we look for horizons outside
some fixed coordinate radius $r_0$. If $r_0$ is a timelike coordinate in
the Reissner-Nordstrom space-time the singularity in the Born-Infeld-Einstein
space-time is spacelike and if $r_0$ is timelike the singularity is spacelike.
  
This implies that charged black holes 
have either two horizons and a timelike singularity or one horizon with a spacelike, timelike, or null singularity.
\section*{Cosmological space-times}
If $m(r)>\frac{1}{2}r$ for $r>\sqrt{a
\tilde{Z}}$, then $r$ is
a timelike coordinate and $t$ is a spacelike coordinate \cite{Vo3}. Relabeling
$r$ and $t$ and denoting the spacelike variable by $x$ gives
\begin{equation}
ds^2=-\left[\frac{2m(t)}{t}-1\right]^{-1}dt^2+\left[\frac{2m(t)}{t}-1\right]
dx^2+t^2d\Omega^2\;,
\end{equation}
\begin{equation}
D=\frac{Q}{t^2}\;dx\wedge dt\;\;\;\; and \;\;\;\; B=P\sin\theta\;
d\theta\wedge d\phi\; ,
\end{equation}
\begin{equation}
\frac{dm(t)}{dt}=\frac{1}{2}\Lambda t^2 +4\pi t^2H(t)\; ,
\end{equation}
and 
\begin{equation}
H(t)=\frac{1}{a^2}(1+\tilde{\lambda})^{-1}\left[1-\sqrt{1-\frac{a^2\tilde{Z}^2}
{t^4}}\right]\;.
\end{equation}
Note that constant timelike surfaces have the topology $R\times S^2$ and
the two sphere has radius $t$.
The condition $\frac{2m(t)}{t}-1>0$ requires that
\begin{equation}
\frac{\alpha}{t}+\frac{\Lambda}{3}t^2-1+\frac{8\pi}{a^2t}(a\tilde{Z})
^{3/2}(1+\tilde{\lambda})^{-1}
\int^{\frac{t}{\sqrt{a\tilde{Z}}}}_1
\left[u^2-\sqrt{u^4-1}\right]du>0\;,
\label{cosm}
\end{equation}
where $\alpha$ is a constant. Since the integral is positive ($\tilde{\lambda}>-1$)
for all $t>\sqrt{
a\tilde{Z}}$ the inequality will be satisfied if $\alpha>\frac{2}{3}
\Lambda^{-1/2}$, with $\Lambda>0$. This condition ensures that the first
three terms in (\ref{cosm}) are greater than zero on $t\in (0,\infty)$.
Thus, this condition is sufficient but not necessary. By computing
the Ricci tensor, as was done in the previous section, it is easy to show
that the space-time is singular at $t=\sqrt{a\tilde{Z}}$.
  
Note that as $t\rightarrow\sqrt{a\tilde{Z}}$ the radius
of the two sphere approaches $\sqrt{a\tilde{Z}}$ and
\begin{equation}
g_{xx}\rightarrow\frac{\alpha}{\sqrt{a\tilde{Z}}}+
\frac{1}{3}a\Lambda\tilde{Z}-1\; .
\end{equation}
Thus, all the metric components generally are nonzero and finite as the initial singularity is approached. At large $t$ the cosmological constant
term will dominate and the Universe will approach a de Sitter space-time.
\section*{Conclusion}
In this paper I examined black hole and cosmological space-times in Born-Infeld-Einstein
theory with electric and magnetic charges. It was shown that the field equations
can be
written in the form $G_{\mu\nu}=-\kappa T_{\mu\nu}$ for spherically symmetric
space-times. The energy-momentum tensor is not the Born-Infeld energy-momentum
tensor, but can be obtained from Born-Infeld theory by letting $a\rightarrow
ia$, where $a$ is the Born-Infeld parameter. It was shown spherically
symmetric space-times have a curvature singularity at a nonzero
radial coordinate and that, as in the Reissner-Nordstrom space-time, there
are zero, one or two horizons. It was also shown that charged black holes 
have either two horizons and a timelike singularity or one horizon with a spacelike, timelike, or null singularity. Anisotropic cosmological solutions with electric
and magnetic fields were obtained from the spherically symmetric 
solutions. These cosmological space-times have topology $R\times S^2$ and
as the initial singularity is approached all the metric components generally
remain finite and nonzero.
\section*{Acknowledgements}
This work was supported by the Natural Sciences and Engineering Research
Council of Canada.


\begin{thebibliography}{1}
\bibitem{Bo1}
M. Born and L. Infeld, Proc. Roy. Soc. {\bf 144}, 425 (1934)
\bibitem{Po1}
J. Polchinski, String Theory Vol. 1, Cambridge University Press (1998)
\bibitem{Vo1}
D.N. Vollick, Phys. Rev. D{\bf 69}, 064030 (2004) 
\bibitem{Vo2}
D.N. Vollick, Phys. Rev. D{\bf 72}, 084026 (2005)
\bibitem{De1}
S. Deser and G.W. Gibbons, Class. Quant. Grav. 15, L{\bf 35} (1998)
\bibitem{Fe1}
J.A. Feigenbaum, P.G.O. Freund and M. Pigli, Phys. Rev. D{\bf 57}, 4738 (1998)
\bibitem{Fe2}
J.A. Feigenbaum, Phys. Rev. D{\bf 58}, 124023 (1998)
\bibitem{Ho1}
B. Hoffmann, Quart. J. Maths (Oxford){\bf 6}, 149 (1935)
\bibitem{Ho2}
B. Hoffmann and I. Infeld, Phys. Rev. {\bf 51}, 765 (1937)
\bibitem{Pe1}
A. Peres, Phys. Rev. {\bf 122}, 273 (1961) 
\bibitem{Dem1}
M. Demianski, Found. Phys. {\bf 16}, 187 (1986)
\bibitem{Wi1}
D. Wiltshire, Phys. Rev. D{\bf 38}, 2445 (1988)
\bibitem{Ol1}
H. de Oliviera, Class. Quant. Grav. {\bf 11}, 1469 (1994)
\bibitem{Vo3}
D.N. Vollick, Gen. Rel. Grav. {\bf 35}, 1511 (2003)
\end{thebibliography}
\end{document}